\documentclass{revtex4}
\usepackage{amssymb}
\usepackage{amsfonts}
\usepackage{amsmath}

\input{tcilatex}

\begin{document}

\title{Position-dependent mass Lagrangians: nonlocal transformations,
Euler-Lagrange invariance and exact solvability }
\author{Omar Mustafa}
\email{omar.mustafa@emu.edu.tr}
\affiliation{Department of Physics, Eastern Mediterranean University, G. Magusa, north
Cyprus, Mersin 10 - Turkey,\\
Tel.: +90 392 6301378; fax: +90 3692 365 1604.}

\begin{abstract}
\textbf{Abstract:} A general nonlocal point transformation for
position-dependent mass Lagrangians and their mapping into a \emph{"constant
unit-mass"} Lagrangians in the generalized coordinates is introduced. The
conditions on the invariance of the related Euler-Lagrange equations are
reported. The harmonic oscillator linearization of the PDM Euler-Lagrange
equations is discussed through some illustrative examples including harmonic
oscillators, shifted harmonic oscillators, a quadratic nonlinear oscillator,
and a Morse-type oscillator. The Mathews-Lakshmanan nonlinear oscillators
are reproduced and some \emph{"shifted"} Mathews-Lakshmanan nonlinear
oscillators are reported. The mapping of an isotonic nonlinear oscillator
into a PDM deformed isotonic oscillator is also discussed.


\end{abstract}

\maketitle

\section{Introduction}

The position-dependent mass (PDM) concept in quantum mechanics has attracted
researchers' attention ever since the introduction of the PDM von Roos
Hamiltonian \cite{1} (see a sample of references in \cite%
{2,3,4,5,6,7,8,9,10,11,12} and related references cited therein). This
research attention is manifested and inspired not only by the PDM feasible
applicability in various fields of physics (e.g., many-body systems,
semiconductors, quantum dots, quantum liquids, etc.), but also by the
mathematical challenge indulged in the von Roos Hamiltonian. Such a
position-dependent mass setting have invigorated a relatively recent and
rapid research attention on the PDM concept for classical mechanical systems
(cf., e.g., \cite{12,13,14,15,16,17,18,19,20,21,22,23,24,25}) which was
readily introduced by Mathews and Lakshmanan \cite{18} back in 1974. Unlike
the ordering ambiguity that arises in the kinetic energy term of the PDM von
Roos quantum mechanical Hamiltonian, no such ordering ambiguities arise in
the classical mechanical PDM Hamiltonian. Yet, it has been asserted that the
resolution of the ordering ambiguity conflict in the PDM quantum Hamiltonian
may be sought in the classical and quantum mechanical correspondence (cf.,
e.g., \cite{12,17} for more details on this issue).

Very recently \cite{22} we have advocated that the equivalence between the
Euler-Lagrange's and Newton's equations of motion is secured through some
"good" invertible coordinate transformation (i.e., $\partial x/\partial
q\neq 0\neq \partial q/\partial x$) and the introduction of the new
PDM-byproducted reaction-type force $R_{PDM}\left( x,\dot{x}\right)
=m^{^{\prime }}\left( x\right) \dot{x}^{2}/2$ into Newton$^{\prime }$s law
of motion (with the overhead dot representing time derivative and the prime
denoting coordinate derivative). Hereby, we have shown that whilst the
quasi-linear momentum is a conserved quantity (i.e., $\Pi \left( x,\dot{x}%
\right) =\Pi _{0}\left( x_{0},\dot{x}_{0}\right) $ and $\dot{\Pi}\left( x,%
\dot{x}\right) =0$) and not the linear momentum (i.e., $p\left( x,\dot{x}%
\right) \neq p_{0}\left( x_{0},\dot{x}_{0}\right) $, and $\dot{p}\left( x,%
\dot{x}\right) \neq 0$), the total energy remains conservative. Therein, we
have conclude that the PDM setting is nothings but a manifestation of some
"good" invertible coordinate transformation that leaves the corresponding
Euler-Lagrange equation invariant. That is, for the PDM-Lagrangian%
\begin{equation}
L=T-V=\frac{1}{2}m\left( x\right) \dot{x}^{2}-V\left( x\right) ,
\end{equation}%
the corresponding Euler-Lagrange equation 
\begin{equation}
\ddot{x}+\frac{1}{2}\frac{m^{^{\prime }}\left( x\right) }{m\left( x\right) }%
\dot{x}^{2}+\frac{1}{m\left( x\right) }\frac{\partial V\left( x\right) }{%
\partial x}=0,
\end{equation}%
is invariant under invertible coordinate transformations. For more details
on this issue the reader may refer to \cite{17,22}. Obviously, moreover,
equation (2) is a quadratic Li\'{e}nard-type differential equation
(quadratic in terms of $\dot{x}^{2}$ in (2)) which serves as a very
interesting model in both physics and mathematics (cf., e.g., the sample of
references \cite{18,19,20,21,22,23,24,25,26,27,28,29,30,31,32,33,34,35,36,37}
and related references cited therein).

In fact, the position-dependent mass concept may very well represent a
position-dependent deformation of the mass. Which would, in turn, manifest
some deformation in the potential force field the mass is moving within and
may inspire nonlocal space-time point transformations. That is, if the
position-dependent mass $m\left( x\right) =m_{\circ }M\left( x\right) $ ($%
m_{\circ }$ is the standard constant mass and is taken as a unit mass
throughout this work) is moving in a harmonic oscillator potential $V\left(
x\right) =m\left( x\right) \omega ^{2}x^{2}/2$, then one may rewrite $%
V\left( x\right) =m_{\circ }\omega ^{2}M\left( x\right)
x^{2}/2\Longrightarrow V\left( u\right) =m_{\circ }\omega ^{2}u^{2}/2;$ $u=%
\sqrt{M\left( x\right) }x$ to retain the standard format for the constant
mass settings. In the process, one may need to use some position-dependent
deformed/rescaled time as well. This is the focal point of the current
methodical proposal.

This paper is organized as follows. In section 2, we introduce a generalized
PDM nonlocal point transformation that maps a PDM-Lagrangian into a \emph{%
"constant unit-mass"} Lagrangian in the generalized coordinates and report
on the conditions that secure the invariance of the related Euler-Lagrange
equations. The harmonic oscillator linearization of the PDM Euler-Lagrange
equations is discussed in section 3. To illustrate our methodical proposal,
we consider (in the same section) some PDM Lagrangians for PDM particles
moving in potential force fields (i) of a harmonic oscillator nature (i.e., $%
V\left( x\right) \sim m\left( x\right) x^{2}$), (ii) of only PDM-dependent
nature (i.e., $V\left( x\right) \sim m\left( x\right) $), (iii) of a shifted
harmonic oscillator nature, (i.e., $V\left( x\right) \sim m\left( x\right)
\left( x+\xi \right) ^{2}$), (iv) of a quadratic nonlinear oscillator nature
(i.e., $V\left( x\right) \sim m\left( x\right) \left( 1+2\lambda x\right)
\left( 1+\lambda x\right) ^{2}$), and (v) of a Morse-type oscillator nature
(i.e., $V\left( x\right) \sim m\left( x\right) \left( 1-e^{-\eta x}\right)
^{2}$). We observe that whilst the Mathews-Lakshmanan nonlinear oscillators 
\cite{18,19,20,21,22,23,24,29,30,31} are reproduced in case (i) and (ii),
some \emph{"shifted"} Mathews-Lakshmanan nonlinear oscillators are obtained
in case (iii) for the same PDM $m\left( x\right) =1/\left( 1\pm \lambda
x^{2}\right) ;$ $\lambda \geq 0$. Moreover, to show that the usage of the
current methodical proposal is not only limited to oscillator linearization,
we discuss (in section 4) the mapping of\ an isotonic nonlinear oscillator
into a PDM deformed isotonic oscillator. We conclude in section 5.

\section{PDM Lagrangians; nonlocal PDM-point transformation and invariance}

Consider a classical particle with a constant \emph{"unit mass"} moving in
the generalized coordinate $q=q\left( x\right) $, a potential force field $%
V(q),$ and a deformed/rescaled time $\tau $. In this case, the Lagrangian
for such a system is given by%
\begin{equation}
L\left( q,\tilde{q},\tau \right) =\frac{1}{2}\tilde{q}^{2}-V(q)\text{ };%
\text{ \ }\tilde{q}=\frac{dq}{d\tau },
\end{equation}%
and the\ corresponding Euler-Lagrange's equation, therefore, reads 
\begin{equation}
\frac{d}{d\tau }\left( \frac{\partial L}{\partial \tilde{q}}\right) -\frac{%
\partial L}{\partial q}=0\Longrightarrow \text{\ }\frac{d^{2}q\left(
x\right) }{d\tau ^{2}}+\frac{\partial V(q)}{\partial q}=0.
\end{equation}%
The introduction of an invertible nonlocal point transformation of the form%
\begin{equation}
q\equiv q\left( x\right) =\int \sqrt{g\left( x\right) }dx\text{ , \ }\tau
=\int f\left( x\right) dt\Longrightarrow \frac{d\tau }{dt}=f\left( x\right)
\neq 0\text{ };\text{ \ }x\equiv x\left( t\right) .
\end{equation}%
could prove to be quite handy in the process. Under such nonlocal
transformation settings, one may, in a straightforward manner, show that%
\begin{equation}
\frac{dq}{d\tau }=\tilde{q}=\frac{\dot{x}\sqrt{g\left( x\right) }}{f\left(
x\right) }=\dot{x}\sqrt{m\left( x\right) }\text{ },\text{ \ \ }\frac{d^{2}q}{%
d\tau ^{2}}=\frac{\sqrt{g\left( x\right) }}{f\left( x\right) ^{2}}\left[ 
\ddot{x}+\frac{1}{2}\left( \frac{g^{^{\prime }}\left( x\right) }{g\left(
x\right) }-2\frac{f^{^{\prime }}\left( x\right) }{f\left( x\right) }\right) 
\dot{x}^{2}\right]
\end{equation}%
and, in turn, equation (4) reads%
\begin{equation}
\left[ \ddot{x}+\frac{1}{2}\left( \frac{g^{^{\prime }}\left( x\right) }{%
g\left( x\right) }-2\frac{f^{^{\prime }}\left( x\right) }{f\left( x\right) }%
\right) \dot{x}^{2}\right] +\frac{f\left( x\right) ^{2}}{g\left( x\right) }%
\frac{\partial V\left( x\right) }{\partial x}=0.
\end{equation}%
Obviously, the comparison between equation (7) and (2) suggests that the
Euler-Lagrange equations of motion (2) and (7) are identical if and only if $%
f\left( x\right) $ and $g\left( x\right) $ satisfy the condition 
\begin{equation}
g\left( x\right) =m\left( x\right) f\left( x\right) ^{2}\Longleftrightarrow 
\frac{g^{^{\prime }}\left( x\right) }{g\left( x\right) }-2\frac{f^{^{\prime
}}\left( x\right) }{f\left( x\right) }=\frac{m^{^{\prime }}\left( x\right) }{%
m\left( x\right) }.
\end{equation}%
One may now conclude that under such conditionally, (8), invertible (i.e.,
the Jacobian determinant $det\left( \partial x_{i}/\partial q_{i}\right)
\neq 0$) nonlocal point transformation, the PDM Euler-Lagrangian equations'
invariance is secured. That is,%
\begin{equation}
L\left( q,\tilde{q},\tau \right) =\frac{1}{2}\tilde{q}^{2}-V(q)%
\Longleftrightarrow \left\{ 
\begin{array}{c}
q\left( x\right) =\int \sqrt{m\left( x\right) }f\left( x\right) dx\medskip
\\ 
\tau =\int f\left( x\right) dt\medskip \\ 
g\left( x\right) =m\left( x\right) f\left( x\right) ^{2}\medskip \\ 
\tilde{q}=\dot{x}\sqrt{m\left( x\right) }%
\end{array}%
\right\} \Longleftrightarrow L\left( x,\dot{x},t\right) =\frac{1}{2}m\left(
x\right) \dot{x}^{2}-V\left( x\right) ,
\end{equation}%
\emph{\ }and hence the Euler-Lagrange equations of motion (4) and (7) are
invariant under the nonlocal point transformation (9), therefore. In fact,
this is a documentation that the Euler-Lagrangian\ equation remains
invariant under some local and nonlocal invertible point transformation
(cf., e.g., Muriel and Romero \cite{32} and Pradeep et al. \cite{33} and
related references cited therein).

At this very point, one should notice that our invertible nonlocal point
transformation (9) is just a subset of the well known \emph{generalized
Sundman transformations} (cf., e.g., \cite{34,35,36,37} and related
references cited therein)%
\begin{equation*}
X=K\left( x,t\right) \text{ , \ \ }dT=N\left( x,t\right) dt\text{ , \ }%
N\left( x,t\right) \frac{\partial K\left( x,t\right) }{\partial x}\neq 0,
\end{equation*}%
that are used to define \emph{Sundman symmetries} (cf., e.g., \cite{32,34,37}%
). Here, 
\begin{equation*}
T=\tau \text{, }N\left( x,t\right) =f\left( x\right) \text{, }X=q\left(
x\right) \text{, }K\left( x,t\right) =\int \sqrt{m\left( x\right) }f\left(
x\right) dx\text{, }N\left( x,t\right) \frac{\partial K\left( x,t\right) }{%
\partial x}=\sqrt{m\left( x\right) }f\left( x\right) ^{2}\neq 0.
\end{equation*}%
The connection between our nonlocal transformation (9) and the \emph{%
generalized Sundman transformation} \cite{32} is therefore clear. Of course,
the nonlocality is an obvious manifestation of the nonlocal term%
\begin{equation*}
T=\int N\left( x,t\right) dt.
\end{equation*}%
Nevertheless, such nonlocal transformations are usually used in the
linearization of a class of nonlinear ordinary differential equations (ODEs)
to transform them into solvable linear ODEs (an interesting issue for
nonlinear physical problems). However, we shall use our nonlacal point
transformation (9) beyond the linearization of ODEs. Namely, and in a more
simplistic language, our nonlocal point transformation (9) also offers some
mapping between Euler-Lagrange equations (4) and (7), where the solution of
one of them (hence, $L\left( q,\tilde{q},\tau \right) $ or $L\left( x,\dot{x}%
,t\right) $ is the \emph{reference Lagrangian}) would lead to the solution
of the other (hence, $L\left( x,\dot{x},t\right) $ or $L\left( q,\tilde{q}%
,\tau \right) $ becomes the \emph{target Lagrangian}). This shall be
illustrated in the forthcoming examples/models below.

\section{Oscillator-Linearization of the PDM Euler-Lagrange equation}

We now consider the classical particle with a constant \emph{"unit mass"}
moving in the above conditionally time-rescaled generalized coordinate (3)
under the influence of the potential force field $V(q)=\omega ^{2}q^{2}/2$.
Then, the oscillator Lagrangian 
\begin{equation}
L\left( q,\tilde{q},\tau \right) =\frac{1}{2}\tilde{q}^{2}-\frac{1}{2}\omega
^{2}q^{2},
\end{equation}%
would yield a linear oscillator-type Euler-Lagrange equation in the form of%
\begin{equation}
\frac{d^{2}q\left( x\right) }{d\tau ^{2}}+\omega ^{2}q=0.
\end{equation}%
With some suitable initial conditions ($q\left( 0\right) =A$, and $\tilde{q}%
\left( 0\right) =0$, say), it admits the periodic solution%
\begin{equation*}
q\left( x\right) =A\cos \left( \omega \tau +\phi \right) .
\end{equation*}%
In this case, equations (4) and (6) would imply%
\begin{equation}
\ddot{x}+\frac{1}{2}\frac{m^{^{\prime }}\left( x\right) }{m\left( x\right) }%
\dot{x}^{2}+\omega ^{2}\frac{f\left( x\right) }{\sqrt{m\left( x\right) }}%
q\left( x\right) =0.
\end{equation}%
Which, when compared with (2) results in the linear-oscillator mapping
condition%
\begin{equation}
\frac{1}{m\left( x\right) }\frac{\partial V\left( x\right) }{\partial x}%
=\omega ^{2}\frac{f\left( x\right) }{\sqrt{m\left( x\right) }}q\left(
x\right) \Longleftrightarrow q\left( x\right) =\frac{1}{\omega ^{2}\sqrt{%
m\left( x\right) }f\left( x\right) }\frac{\partial V\left( x\right) }{%
\partial x}
\end{equation}%
Therefore, the mapping between the conditionally time-rescaled generalized
coordinate system and the Cartesian one-dimensional $x$-coordinate is now
clear. We illustrate the applicability of the above oscillator-linearization
procedure through the following examples.

\subsection{Mathews-Lakshmanan PDM-nonlinear Oscillators I; an $f\left(
x\right) =m\left( x\right) $ case}

Let us consider the PDM particle moving in the harmonic oscillator force
field $V\left( x\right) =m\left( x\right) \omega ^{2}x^{2}/2$ with the
corresponding Lagrangian%
\begin{equation}
L\left( x,\dot{x},t\right) =\frac{1}{2}m\left( x\right) \dot{x}^{2}-\frac{1}{%
2}m\left( x\right) \omega ^{2}x^{2}\text{.}
\end{equation}%
Then, the linear-oscillator mapping condition (13) would suggest%
\begin{equation}
q\left( x\right) =x\sqrt{m\left( x\right) }=\int \sqrt{m\left( x\right) }%
f\left( x\right) dx,
\end{equation}%
provided that 
\begin{equation}
f\left( x\right) =\left( 1+\frac{1}{2}\frac{m^{^{\prime }}\left( x\right) }{%
m\left( x\right) }x\right) .
\end{equation}%
Moreover, the assumption that $f\left( x\right) =m\left( x\right) $ would
result in a specific PDM-function. That is,%
\begin{equation}
f\left( x\right) =1+\frac{1}{2}\frac{m^{^{\prime }}\left( x\right) }{m\left(
x\right) }x=m\left( x\right) \Longleftrightarrow f\left( x\right) =m\left(
x\right) =\frac{1}{1\pm \lambda x^{2}}\text{ };\text{ \ }\lambda \geq 0.
\end{equation}%
Therefore, the nonlocal transformation (9) now (with $f\left( x\right)
=m\left( x\right) $) reads%
\begin{equation}
L\left( q,\tilde{q},\tau \right) =\frac{1}{2}\tilde{q}^{2}-\frac{1}{2}\omega
^{2}q^{2}\Longleftrightarrow \left\{ 
\begin{array}{c}
q\left( x\right) =x\sqrt{m\left( x\right) }\medskip \\ 
\tilde{q}=\dot{x}\sqrt{m\left( x\right) }\medskip \\ 
d\tau /dt=m\left( x\right) \medskip \\ 
g\left( x\right) =m\left( x\right) ^{3}\medskip%
\end{array}%
\right\} \Longleftrightarrow L\left( x,\dot{x},t\right) =\frac{1}{2}\frac{%
\left( \dot{x}^{2}-\omega ^{2}x^{2}\right) }{1\pm \lambda x^{2}};\text{ \ }%
\lambda \geq 0,
\end{equation}%
and, in turn, yields the Mathews-Lakshmanan oscillators' equations \cite%
{23,24,25}%
\begin{equation}
\ddot{x}\mp \frac{\lambda x}{1\pm \lambda x^{2}}\dot{x}^{2}+\frac{\omega
^{2}x}{1\pm \lambda x^{2}}=0.
\end{equation}%
with the solutions%
\begin{equation}
x\left( t\right) =A\cos \left( \Omega t+\phi \right) \,;\text{ \ }\Omega
^{2}=\frac{\omega ^{2}}{1\pm \lambda A^{2}}.
\end{equation}%
Hereby, one may mention that the PDM-quantum mechanical versions of such
nonlinear oscillators are studied by Lakshmanan and Chandrasekar \cite{24}
and by Cari\~{n}ena et al. \cite{30,31}, where the transition from the
PDM-classical system to the PDM-quantum one is analyzed through some
differential geometric factorization recipe. For more details on this issue
the reader may refer to \cite{24,30,31} and related references cited therein.

\subsection{Mathews-Lakshmanan PDM-nonlinear Oscillators II; an $f\left(
x\right) =\protect\beta m^{\prime }\left( x\right) /2m\left( x\right) $ case}

Consider the PDM particle moving in a force field of the form $V\left(
x\right) =\beta ^{2}\omega ^{2}m\left( x\right) /2$, where $\beta $ is a
non-zero constant introduced for the convenience of calculations. Then the
corresponding Lagrangian is given by%
\begin{equation}
L\left( x,\dot{x},t\right) =\frac{1}{2}m\left( x\right) \dot{x}^{2}-\frac{%
\omega ^{2}}{2}\beta ^{2}m\left( x\right) .
\end{equation}%
With the choice that%
\begin{equation}
f\left( x\right) =\beta \frac{m^{\prime }\left( x\right) }{2m\left( x\right) 
}\Longrightarrow q\left( x\right) =\beta \int \frac{1}{2}\frac{m^{\prime
}\left( x\right) }{\sqrt{m\left( x\right) }}dx=\beta \sqrt{m\left( x\right) }%
\Longrightarrow m\left( x\right) =\exp \left( \frac{2}{\beta }\int f\left(
x\right) dx\right) ,
\end{equation}%
one may immediately show that the nonlocal transformation (9) reads%
\begin{equation}
L\left( q,\tilde{q},\tau \right) =\frac{1}{2}\tilde{q}^{2}-\frac{1}{2}\omega
^{2}q^{2}\Longleftrightarrow \left\{ 
\begin{array}{c}
q\left( x\right) =\beta \sqrt{m\left( x\right) }\medskip \\ 
\tilde{q}=\dot{x}\sqrt{m\left( x\right) }\medskip \\ 
d\tau /dt=\beta m^{\prime }\left( x\right) /2m\left( x\right) \medskip \\ 
g\left( x\right) =\beta m^{\prime }\left( x\right) ^{2}/4m\left( x\right)
\medskip%
\end{array}%
\right\} \Longleftrightarrow L\left( x,\dot{x},t\right) =\frac{1}{2}m\left(
x\right) \left( \dot{x}^{2}-\beta ^{2}\omega ^{2}\right) ,
\end{equation}%
to imply the Euler-Lagrange equation%
\begin{equation}
\ddot{x}+\frac{1}{2}\frac{m^{^{\prime }}\left( x\right) }{m\left( x\right) }%
\dot{x}^{2}+\beta ^{2}\omega ^{2}\frac{m^{^{\prime }}\left( x\right) }{%
2m\left( x\right) }=0.
\end{equation}%
In this case, moreover, the Mathews-Lakshmanan oscillators' equations \cite%
{23,24,25}%
\begin{equation}
\ddot{x}\mp \frac{\lambda x}{1\pm \lambda x^{2}}\dot{x}^{2}+\frac{\omega
^{2}x}{1\pm \lambda x^{2}}=0\,;\,\lambda =\mp \frac{1}{\beta ^{2}};\,\lambda
\geq 0,
\end{equation}%
are obtained using%
\begin{equation}
m\left( x\right) =\frac{1}{1\pm \lambda x^{2}}.
\end{equation}%
and admit solutions in the form of%
\begin{equation}
x\left( t\right) =A\cos \left( \Omega t+\phi \right) \,;\text{ \ }\Omega
^{2}=\frac{\omega ^{2}}{1\pm \lambda A^{2}}.
\end{equation}

Nevertheless, one should also notice that the Mathews-Lakshmanan
oscillators' equations are obtained by direct substitutions of the set of
four potential force fields%
\begin{equation}
V\left( x\right) =\left\{ 
\begin{tabular}{cc}
$-m\left( x\right) \omega ^{2}\mathstrut /2\lambda ;\medskip $ & $\text{for }%
m\left( x\right) =1/\left( 1+\lambda x^{2}\right) $ \\ 
$+m\left( x\right) \omega ^{2}\mathstrut /2\lambda ;\medskip $ & $\text{for }%
m\left( x\right) =1/\left( 1-\lambda x^{2}\right) $ \\ 
$\frac{1}{2}m\left( x\right) \omega ^{2}x^{2};$ & $\text{for }m\left(
x\right) =1/\left( 1\pm \lambda x^{2}\right) $%
\end{tabular}%
\right.
\end{equation}%
in (2) (each potential at a time, of course). This is very much related to
the nature of the given PDM function.

\subsection{Shifted Mathews-Lakshmanan PDM-nonlinear Oscillators III; an $%
f\left( x\right) =m\left( x\right) $ case}

Let us consider the PDM particle moving in the shifted harmonic oscillator
force field $V\left( x\right) =m\left( x\right) \omega ^{2}\left( x+\xi
\right) ^{2}/2$. Then the corresponding Lagrangian is given by%
\begin{equation}
L\left( x,\dot{x},t\right) =\frac{1}{2}m\left( x\right) \dot{x}^{2}-\frac{1}{%
2}m\left( x\right) \omega ^{2}\left( x+\xi \right) ^{2}.
\end{equation}%
With the choice that%
\begin{equation}
q\left( x\right) =\left( x+\xi \right) \sqrt{m\left( x\right) }%
\Longrightarrow \frac{dq\left( x\right) }{dx}=\sqrt{m\left( x\right) }%
f\left( x\right) \Longrightarrow f\left( x\right) =\left( 1+\frac{1}{2}\frac{%
m^{^{\prime }}\left( x\right) }{m\left( x\right) }\left( x+\xi \right)
\right) ,
\end{equation}%
the assumption $f\left( x\right) =m\left( x\right) $ would yield the
specific PDM-function of the form%
\begin{equation}
f\left( x\right) =m\left( x\right) =\frac{1}{1\pm \lambda \left( x+\xi
\right) ^{2}}\text{ };\text{ \ }\lambda \geq 0.
\end{equation}%
Therefore, the nonlocal transformation (9) now reads%
\begin{equation}
L\left( q,\tilde{q},\tau \right) =\frac{1}{2}\tilde{q}^{2}-\frac{1}{2}\omega
^{2}q^{2}\Longleftrightarrow \left\{ 
\begin{array}{c}
q\left( x\right) =\left( x+\xi \right) \sqrt{m\left( x\right) }\medskip \\ 
\tilde{q}=\dot{x}\sqrt{m\left( x\right) }\medskip \\ 
d\tau /dt=m\left( x\right) \medskip \\ 
g\left( x\right) =m\left( x\right) ^{3}\medskip%
\end{array}%
\right\} \Longleftrightarrow L\left( x,\dot{x},t\right) =\frac{1}{2}\frac{%
\left[ \dot{x}^{2}-\omega ^{2}\left( x+\xi \right) ^{2}\right] }{1\pm
\lambda x^{2}};\text{ \ }\lambda \geq 0,
\end{equation}%
to imply the shifted Mathews-Lakshmanan oscillators' equations%
\begin{equation}
\ddot{x}\mp \frac{\lambda \left( x+\xi \right) }{1\pm \lambda \left( x+\xi
\right) ^{2}}\dot{x}^{2}+\frac{\omega ^{2}\left( x+\xi \right) }{1\pm
\lambda \left( x+\xi \right) ^{2}}=0.
\end{equation}%
with the solutions%
\begin{equation}
x\left( t\right) =A\cos \left( \Omega t+\phi \right) -\xi \,;\text{ \ }%
\Omega ^{2}=\frac{\omega ^{2}}{1\pm \lambda A^{2}}.
\end{equation}

Yet, one can follow (step-by-step) the nonlocal transformation recipe in
(23) to show that if the PDM in (31) is moving in the potential force field $%
V\left( x\right) =\beta ^{2}\omega ^{2}m\left( x\right) /2$, it would admit
similar Euler-Lagrange dynamical equations of motion as those in (33) and
exactly follow the same trajectory as that in (34). Under such PDM settings,
one would observe that the shifted Mathews-Lakshmanan oscillators' equations
(33) are obtained by direct substitutions of the set of four potential force
fields%
\begin{equation}
V\left( x\right) =\left\{ 
\begin{tabular}{cc}
$-m\left( x\right) \omega ^{2}\mathstrut /2\lambda ;\medskip $ & $\text{for }%
m\left( x\right) =1/\left( 1+\lambda \left( x+\xi \right) ^{2}\right) $ \\ 
$+m\left( x\right) \omega ^{2}\mathstrut /2\lambda ;\medskip $ & $\text{for }%
m\left( x\right) =1/\left( 1-\lambda \left( x+\xi \right) ^{2}\right) $ \\ 
$\frac{1}{2}m\left( x\right) \omega ^{2}\left( x+\xi \right) ^{2};$ & $\text{%
for }m\left( x\right) =1/\left( 1\pm \lambda \left( x+\xi \right)
^{2}\right) $%
\end{tabular}%
\right.
\end{equation}%
in (2) (each potential at a time, of course).

\subsection{A Quadratic non-linear PDM Oscillator; an $f\left( x\right) =1$
case}

Consider a PDM particle moving in the potential force field 
\begin{equation}
V\left( x\right) =-\frac{\alpha ^{2}}{2\lambda ^{2}}m\left( x\right) \left(
1+2\lambda x\right) \left( 1+\lambda x\right) ^{2}
\end{equation}%
with the corresponding Lagrangian%
\begin{equation}
L\left( x,\dot{x},t\right) =\frac{1}{2}m\left( x\right) \dot{x}^{2}+\frac{%
\alpha ^{2}}{2\lambda ^{2}}m\left( x\right) \left( 1+2\lambda x\right)
\left( 1+\lambda x\right) ^{2},
\end{equation}%
Let us now defined a point transformation (with $f\left( x\right) =1$ ) of
the form%
\begin{equation}
q\left( x\right) =xm\left( x\right) ^{1/4}\medskip =\int \sqrt{m\left(
x\right) }f\left( x\right) dx.
\end{equation}%
In this case, one obtains%
\begin{equation}
1+\frac{1}{4}\frac{m^{^{\prime }}\left( x\right) }{m\left( x\right) }%
x=m\left( x\right) ^{1/4}\Longrightarrow m\left( x\right) =\frac{1}{\left(
1+\lambda x\right) ^{4}}\Longrightarrow q\left( x\right) =\frac{x}{1+\lambda
x}.
\end{equation}%
Under such settings, the nonlocal transformation (9) reads%
\begin{equation}
L\left( q,\tilde{q},\tau \right) =\frac{1}{2}\tilde{q}^{2}-\frac{1}{2}\omega
^{2}q^{2}\Longleftrightarrow \left\{ 
\begin{array}{c}
q\left( x\right) =xm\left( x\right) ^{1/4} \\ 
\tilde{q}=\dot{x}\sqrt{m\left( x\right) }\medskip \\ 
d\tau /dt=1\medskip \\ 
g\left( x\right) =m\left( x\right) \medskip%
\end{array}%
\right\} \Longleftrightarrow L\left( x,\dot{x},t\right) =\frac{1}{2}m\left(
x\right) \left( \dot{x}^{2}+\frac{\alpha ^{2}}{\lambda ^{2}}\left(
1+2\lambda x\right) \left( 1+\lambda x\right) ^{2}\right) ,
\end{equation}%
and yields the non-linear \ quadratic oscillator equation \cite{23,24}%
\begin{equation}
\ddot{x}-\frac{2\lambda }{1+\lambda x}\dot{x}^{2}+\alpha ^{2}x\left(
1+\lambda x\right) =0\,;\text{ \ }\alpha ^{2}=\omega ^{2},
\end{equation}%
that admits a solution of the form%
\begin{equation}
x\left( t\right) =\frac{A\cos \left( \alpha t+\phi \right) }{1-\lambda A\cos
\left( \alpha t+\phi \right) };\text{ \ }0\leq A<\frac{1}{\lambda }.
\end{equation}

\subsection{A Morse-oscillator; an $f\left( x\right) =\protect\eta $ case}

Consider a PDM particle moving in a Morse-type oscillator force field with
the corresponding Lagrangian%
\begin{equation}
L\left( x,\dot{x},t\right) =\frac{1}{2}m\left( x\right) \dot{x}^{2}-V\left(
x\right) \text{; \ }V\left( x\right) =\frac{1}{2}m\left( x\right) \alpha
^{2}\left( 1-e^{-\eta x}\right) ^{2}
\end{equation}%
Let us defined a nonlocal transformation of the form%
\begin{equation}
q\left( x\right) =\sqrt{m\left( x\right) }-1\medskip =\int \sqrt{m\left(
x\right) }f\left( x\right) dx.
\end{equation}%
Then one obtains%
\begin{equation}
\sqrt{m\left( x\right) }f\left( x\right) =\frac{m^{^{\prime }}\left(
x\right) }{2\sqrt{m\left( x\right) }}\Longrightarrow f\left( x\right) =\frac{%
m^{^{\prime }}\left( x\right) }{2m\left( x\right) }=\eta \Longrightarrow
m\left( x\right) =\exp \left( 2\eta x\right) \Longrightarrow \tau =\eta t.
\end{equation}%
Under such settings, the nonlocal transformation (9) reads%
\begin{equation}
L\left( q,\tilde{q},\tau \right) =\frac{1}{2}\tilde{q}^{2}-\frac{1}{2}\omega
^{2}q^{2}\Longleftrightarrow \left\{ 
\begin{array}{c}
q\left( x\right) =\sqrt{m\left( x\right) }-1\medskip \\ 
\tilde{q}=\dot{x}\sqrt{m\left( x\right) }\medskip \\ 
d\tau /dt=\eta \medskip \\ 
g\left( x\right) =\eta ^{2}m\left( x\right) \medskip%
\end{array}%
\right\} \Longleftrightarrow L\left( x,\dot{x},t\right) =\frac{1}{2}m\left(
x\right) \left( \dot{x}^{2}-\alpha ^{2}\left( 1-e^{-\eta x}\right)
^{2}\right) ,
\end{equation}%
and, in turn, implies the Morse-type PDM oscillator's equation \cite{23,24} 
\begin{equation}
\ddot{x}+\eta \dot{x}^{2}+\frac{\alpha ^{2}}{\eta }\left( 1-e^{-\eta
x}\right) =0;\text{ \ }\alpha ^{2}=\omega ^{2}\eta ^{2}
\end{equation}%
with the solution%
\begin{equation}
x\left( t\right) =\frac{1}{\eta }\ln \left( 1+A\cos \left( \alpha t+\phi
\right) \right) \,;0\leq A<1.
\end{equation}

\section{Mapping an isotonic nonlinear-oscillator into a PDM-deformed
isotonic nonlinear-oscillator}

In this section we consider a \emph{unit mass} particle moving in an
isotonic oscillator force field in the above conditionally time-rescaled
generalized coordinate system. Then the corresponding isotonic oscillator
Lagrangian 
\begin{equation}
L\left( q,\tilde{q},\tau \right) =\frac{1}{2}\tilde{q}^{2}-V\left( q\right)
;\,\,V\left( q\right) =\frac{1}{2}\omega ^{2}q^{2}+\frac{\beta }{q^{2}},
\end{equation}%
would imply the Euler-Lagrange equation 
\begin{equation}
\frac{d^{2}q\left( x\right) }{d\tau ^{2}}+\omega ^{2}q-\frac{2\beta }{q^{3}}%
=0,
\end{equation}%
which is known as the Ermakov-Pinney's equation. It admits a general
solution \cite{29} of the form 
\begin{equation}
q=\frac{1}{\omega A}\sqrt{\left( \omega ^{2}A^{4}-2\beta \right) \sin
^{2}\left( \omega \tau +\delta \right) +2\beta }.
\end{equation}

This model follows the conditional nonlocal transformation used for the
Mathews-Lakshmanan PDM-Oscillators I (i.e., the condition that $f\left(
x\right) =m\left( x\right) =1/\left( 1\pm \lambda x^{2}\right) ;\,$\ $%
\lambda \geq 0$). In this case, the nonlocal transformation (9) reads 
\begin{equation}
L\left( q,\tilde{q},\tau \right) =\frac{1}{2}\tilde{q}^{2}-\frac{1}{2}\omega
^{2}q^{2}-\frac{\beta }{q^{2}}\Longleftrightarrow \left\{ 
\begin{array}{c}
q\left( x\right) =x\sqrt{m\left( x\right) }\medskip \\ 
\tilde{q}=\dot{x}\sqrt{m\left( x\right) }\medskip \\ 
d\tau /dt=m\left( x\right) \medskip \\ 
g\left( x\right) =m\left( x\right) ^{3}\medskip%
\end{array}%
\right\} \Longleftrightarrow L\left( x,\dot{x},t\right) =\frac{1}{2}\frac{%
\left( \dot{x}^{2}+\omega ^{2}x^{2}\right) }{1\pm \lambda x^{2}}-\frac{\beta
\left( 1\pm \lambda x^{2}\right) }{x^{2}},
\end{equation}%
and satisfies the Euler-Lagrange equation for a PDM-deformed isotonic
nonlinear-oscillator%
\begin{equation}
\ddot{x}\mp \frac{\lambda x}{1\pm \lambda x^{2}}\dot{x}^{2}+\frac{\omega
^{2}x}{1\pm \lambda x^{2}}-\frac{2\beta }{x^{3}}\left( 1\pm \lambda
x^{2}\right) =0\text{ };\text{ \ }\alpha =\mp \omega ^{2}.
\end{equation}%
The general solution of which is given by%
\begin{equation}
x\left( t\right) =\frac{1}{\Omega A}\sqrt{\left( \Omega ^{2}A^{4}-2\beta
\right) \sin ^{2}\left( \Omega t+\delta \right) +2\beta };\text{ }\omega
^{2}=\left( 1\pm \lambda A^{2}\right) \left( \Omega ^{2}\pm \frac{2\lambda
\beta }{A^{2}}\right) .
\end{equation}%
Hereby, we observe that our nonlocal point transformation (9) has offered
(in addition to the nonlinear-oscillators' linearizations discussed above) a
mapping (52) of a \emph{unit mass} isotonic nonlinear-oscillator into a
PDM-deformed isotonic nonlinear-oscillator. Hence, $L\left( q,\tilde{q},\tau
\right) $ plays the role as a \emph{reference Lagrangian} and $L\left( x,%
\dot{x},t\right) $ as a \emph{target Lagrangian} in (52).

\section{Concluding Remarks}

In this work, we have introduced a general nonlocal point transformation for
PDM Lagrangians and their mapping into a \emph{"constant unit-mass"}
Lagrangians in the generalized coordinates. The conditions on the invariance
of the related Euler-Lagrange equations are also reported. The harmonic
oscillator linearization of the PDM Euler-Lagrange equations is discussed
through some illustrative examples including harmonic oscillators, shifted
harmonic oscillators, a quadratic nonlinear oscillator, and a Morse-type
oscillator. The Mathews-Lakshmanan nonlinear oscillators are reproduced and
some \emph{"shifted"} Mathews-Lakshmanan nonlinear oscillators are reported.
We have also discussed the mapping of an isotonic nonlinear oscillator into
a PDM deformed isotonic oscillator. In the light of the experiment above our
observations are in order.

In connection with the Mathews-Lakshmanan nonlinear oscillators I and II, we
observe that the PDM-function (26) subjected to move in the set of four
potential force fields (28) (One at a time) admits/feels exactly similar
dynamical effects as documented in the corresponding Euler-Lagrange
equations of motion (19) and (25) and follows exactly similar trajectories.
This tendency of similar dynamics, similar trajectories and similar total
energies%
\begin{equation}
E=\frac{1}{2}\omega ^{2}\frac{A^{2}}{1\pm \lambda A^{2}}.
\end{equation}%
is attributed to the nature of the PDM-functional settings used. Similar
dynamics, trajectories and total energies%
\begin{equation}
E=\frac{1}{2}\omega ^{2}\frac{\left( A-\xi \right) ^{2}}{1\pm \lambda \left(
A-\xi \right) ^{2}}.
\end{equation}%
trends are also observed feasible for the shifted Mathews-Lakshmanan
nonlinear oscillators III.

The scope of the applicability of the current methodical proposal extends
beyond the nonlinear-oscillators' linearizations of the PDM Euler-Lagrange
equations (documented through the illustrative examples in section III
above) into the extraction of exact solutions of more complicated dynamical
problems. The mapping of the isotonic nonlinear oscillator in the
generalized coordinates (i.e., \emph{reference/target-Lagrangian}) into a
PDM-deformed isotonic oscillator on the $x$-coordinate (\emph{%
target/reference-Lagrangian}) was just one of such exact-solution
extractions through the nonlocal transformation (9). The extension of the
applicability of our invertible nonlocal transformation (9) may include more
than one dimensional classical systems as well.

\end{document}